\documentclass[aps,showpacs,preprintnumbers,amsmath,prd,amssymb,superscriptaddress]{revtex4}
\usepackage{amssymb}
\usepackage{CJK}
\usepackage{amsmath,amssymb,graphicx,bm}

\begin{document}
\begin{CJK*}{GBK}{}

\title{Spatial Ricci scalar dark energy model}

\author{Rong-Jia Yang}
\email{yangrj08@gmail.com}
 \affiliation{College of Physical Science and Technology,
Hebei University, Baoding 071002, China}

\author{Zong-Hong Zhu}
\affiliation{Department of Astronomy, Beijing Normal University,
Beijing 100875, China}

\author{Fengquan Wu}
\affiliation{The National Astronomical Observatories, Chinese
Academy of Science, Beijing 100012, China}
\date{\today}

\begin{abstract}
Inspired by holographic principle, we suggest that the density of
dark energy is proportional to the spatial Ricci scalar curvature
(SRDE). Such model is phenomenologically viable. The best fit values
of its parameters at $68\%$ confidence level are found to be:
$\Omega_{\rm m0}=0.259\pm0.016$ and $\alpha=0.261\pm0.0122$,
constrained from the Union+CFA3 sample of 397 SNIa and the BAO
measurement. We find the equation of state of SRDE crosses $-1$ at
$z\simeq-0.14$. The present values of the deceleration parameter
$q(z)$ for SRDE is found to be $q_{z=0}\sim -0.85$. The phase
transition from deceleration to acceleration of the Universe for
SRDE occurs at the redshift $z_{q=0}\sim 0.4$. After studying on the
perturbation of each component of the Universe, we show that the
matter power spectra and cosmic microwave background temperature
anisotropy is slightly affected by SRDE, compared with $\Lambda$CDM.
\end{abstract}

\makeatletter
\def\@pacs@name{PACS: 95.36.+x, 98.80.-k, 98.80.Es}
\makeatother

\maketitle
\end{CJK*}

\section{Introduction}
In the last decade a convergence of independent cosmological
observations suggested that the Universe is experiencing accelerated
expansion. An unknown energy component, dubbed as dark energy, may
be responsible for this phenomenon. Numerous dynamical dark energy
models have been proposed in the literature, such as quintessence
\cite{quint,Zlatev}, phantom \cite{phantom,Carroll2003}, $k$-essence \cite{k,Gao2010,Yang2009,Yang2008a,yang10,yang10a}, tachyon
\cite{tachyonic,Bagla2003,Keresztes,yang}, DGP
\cite{Dvali,zhang}, Chaplygin gas \cite{Chaplygin,Bento2002,Gorini05}, ect.
However, the simplest and most theoretically appealing candidate of
dark energy is the vacuum energy (or the cosmological constant
$\Lambda$) with a constant equation of state (EoS) parameter $w=-1$.
This scenario is in general agreement with the current astronomical
observations, but has difficulties to reconcile the small
observational value of dark energy density with estimates from
quantum field theories; this is the cosmological constant problem
(see e. g. \cite{weinberg,Sahni2000,Carroll2001,Copeland2006}). Recently it was shown that $\Lambda$CDM model may also suffer from an age problem \cite{Yang2010a}.
According to the holographic principle
\cite{holography}, a method may help solve this problem was proposed
that an unknown vacuum energy with density proportional to the
Hubble scale, $\Lambda\propto l^{-2}\sim H^2$, could be present
\cite{horova,Minic2000,Thomas2002}, here $l$ is a characteristic length. Unfortunately,
the EoS for such vacuum energy is zero and the Universe is
decelerating. Alternatively, the particle horizon can be introduced
as the characteristic length $l$. However, the EoS is greater than
$-1/3$ in this case; so it still could not explain the observed
acceleration of the Universe \cite{Fischler1998, Bousso1999,Hsu2004, Li2004}.
In view of this, the future event horizon was proposed as the
characteristic length $l$. This holographic dark energy model and
its interacting version are successful in fitting the current
observations \cite{refHDE,Myung2009,Setare2009,Jamil2009,Mohseni2009,
Guberina07,Elizalde2005,Karwan2008,Medved2009,Nayak2009,Cruz2008,Bisabr2009}. However, this holographic dark energy
model has suffered some serious conceptual problems. Firstly, the
present value of dark energy is determined by the future evolution
of the Universe; this poses a challenge to causality. Secondly, the
density of dark energy is a local quantity, while the future event
horizon is a global concept of space-time. To avoid these
short-comes, another possibility is considered: the characteristic
length $l$ is given by the average radius of Ricci scalar curvature
$R^{-1/2}$, in other words, the density of dark energy is
proportional to the Ricci scalar curvature, $\rho_{\rm X}\propto R$
(hereafter RDE for short) \cite{Gao}. Recent studies on this model
see e. g. \cite{Granda,Kim08,Chen2009,Saridakis2008}. However, as holographic principle asserting
\cite{holography}, the number of possible states of a region of
space is the same as that of binary degrees of freedom distributed
on the boundary of the region. To reflect holographic principle more
properly, we propose that the density of dark energy is proportional
to the spatial Ricci scalar curvature, $\rho_{\rm X}\propto R_{\rm
s}$ (hereafter SRDE for short), instead of Ricci scalar curvature.
We find SRDE not only has the same properties, but also has
properties which RDE doesn't have.

This paper is organized as follows. In section II, we describe the
SRDE model. Observations constraints and cosmic expansion will be
discussed in section III. We discuss structure formation in section
IV. Finally, conclusions and discussions are made in section V.

\section{Spatial Ricci scalar dark energy model}
We consider the homogeneous and isotropic
Friedmann-Robertson-Walker-Lema\^{i}tre (FRWL) metric with the scale
factor $a(t)$
\begin{eqnarray}
\label{frwmet}
ds^2=-dt^2+a^2(t)\left[\frac{dr^2}{1-Kr^2}+r^2(d\theta^2+\sin^2\theta
d\phi^2)\right],
\end{eqnarray}
where the spatial curvature constant $K=+1$, 0, and $-1$ correspond
to a closed, flat and open Universe, respectively. We use the units
$c=G=1$ throughout this study. The Friedmann equations reads
\begin{eqnarray}
\label{F1} H^2&=&\frac{8\pi}{3}\sum_{i}\rho_{i}-\frac{K}{a^2}, \\
\label{F2}\dot{H}&=&-4\pi\sum_{i}(\rho_{i}+p_{i})+\frac{K}{a^2},
\end{eqnarray}
where $H\equiv \dot{a}/a$ is the Hubble parameter, the dot denotes
the derivative with respect to the cosmic time $t$, and the
summation runs over the radiation, nonrelativistic matter, and other
components. The conservation equation of the $i$th component of
energy takes the form
\begin{eqnarray}
\label{F3} \dot{\rho_{i}}+3H(\rho_{i}+p_{i})=0.
\end{eqnarray}

The spatial Ricci scalar curvature is given by
\begin{eqnarray}
\label{Ricci} R_{\rm
s}=-3\left[\frac{\ddot{a}}{a}+2\left(\frac{\dot{a}}{a}\right)^2+\frac{2K}{a^2}\right]=-3\left(\dot{H}+3H^2+\frac{2K}{a^2}\right).
\end{eqnarray}
We assume the density of dark energy component is proportional to
the spatial Ricci scalar curvature inspired by the holographic
principle
\begin{eqnarray}
\label{de} \rho_{\rm X}=-\frac{\alpha}{8\pi} R_{\rm
s}=\frac{3\alpha}{8\pi}\left(\dot{H}+3H^2+\frac{2K}{a^2}\right),
\end{eqnarray}
where $\alpha$ is a constant to be determined. The factor $1/8\pi$
is for convenience in the following calculations. Setting $x=\ln a$,
we can rewrite the Friedmann equation (\ref{F1}) as
\begin{eqnarray}
\label{Fn}
H^2&=&\frac{8\pi}{3}\left[(2\alpha-1)\frac{3K}{8\pi}e^{-2x}+\rho_{\rm
m0}e^{-3x}+\rho_{\rm
r0}e^{-4x}\right]\nonumber\\
&+&\alpha\left(\frac{1}{2}\frac{dH^2}{dx}+3H^2\right),
\end{eqnarray}
where the $\rho_{\rm m0}$ and $\rho_{\rm r0}$ are the contributions
of nonrelativistic matter and radiation (here and thereafter the
subscript 0 denotes the value at the present time), respectively. By
introducing the scaled Hubble expansion rate $E\equiv H/H_0$, the
above Friedmann equation becomes
\begin{eqnarray}
\label{Eq} E^2&=&(2\alpha-1)\Omega_{\rm K0}e^{-2x}+\Omega_{\rm
m0}e^{-3x}+\Omega_{\rm
r0}e^{-4x}\nonumber\\
&+&\alpha\left(\frac{1}{2}\frac{dE^2}{dx}+3E^2\right),
\end{eqnarray}
where $\Omega_{\rm K0}$, $\Omega_{\rm m0}$, and $\Omega_{\rm r0}$
are the dimensionless density parameters of the curvature,
nonrelativistic matter, and radiation in the present Universe.
Solving Eq. (\ref{Fn1}), we obtain
\begin{eqnarray}
\label{Fn1} E^2&=&-\Omega_{\rm K0}e^{-2x}+\Omega_{\rm
m0}e^{-3x}+\Omega_{\rm
r0}e^{-4x}\nonumber\\&+&\frac{3\alpha\Omega_{\rm
m0}}{2-3\alpha}e^{-3x}+\frac{\alpha\Omega_{\rm
r0}}{1-\alpha}e^{-4x}+c_0e^{(-6+2/\alpha)x}\nonumber\\
&=&-\Omega_{\rm K0}a^{-2}+\Omega_{\rm m0}a^{-3}+\Omega_{\rm
r0}a^{-4}\nonumber\\&+&\frac{3\alpha\Omega_{\rm
m0}}{2-3\alpha}a^{-3}+\frac{\alpha\Omega_{\rm
r0}}{1-\alpha}a^{-4}+c_0a^{-6+2/\alpha},
\end{eqnarray}
where $c_0$ is an integration constant, the density of spatial
curvature is defined as $\rho_{\rm K}=3K/8\pi a^2$. From the Eq.
(\ref{Fn1}), we obtain the terms come from dark energy.
\begin{eqnarray}
\label{DE1} \Omega_{\rm X}\equiv \rho_{X}/\rho_{\rm
c0}=\frac{3\alpha\Omega_{\rm
m0}}{2-3\alpha}a^{-3}+\frac{\alpha\Omega_{\rm
r0}}{1-\alpha}a^{-4}+c_0a^{-6+2/\alpha},
\end{eqnarray}
where $\rho_{\rm c0}$ is the present value of critical energy
density. The first term of the right hand of Eq. (\ref{DE1}) evolves
like nonrelativistic matter, the last term increases with decreasing
redshift. While unlike RDE,  SRDE also contain one part evolves like
radiation, this term can be seen as dark radiation like as in brane
dark energy \cite{Shtanov}. It is interesting the dark radiation
also present in Ho\u{r}ava-Lifshitz cosmology \cite{Calcagni,Kiritsis2009}, The
equation of state of dark energy reads
\begin{eqnarray}
\label{DEE} w_{\rm X}&=&-1-\frac{1}{3}\frac{1}{\Omega_{\rm
X}}\frac{d\Omega_{\rm X}}{dx}\nonumber\\
&=& \frac{\frac{\alpha\Omega_{\rm
r0}}{3(1-\alpha)}a^{-4}+(1-\frac{2}{3\alpha})c_0a^{-6+2/\alpha}}{\frac{3\alpha\Omega_{\rm
m0}}{2-3\alpha}a^{-3}+\frac{\alpha\Omega_{\rm
r0}}{1-\alpha}a^{-4}+c_0a^{-6+2/\alpha}}.
\end{eqnarray}
For the case of $K=0$, we obtain $c_0=1-2\Omega_{\rm
m0}/(2-3\alpha)-\Omega_{\rm r0}/(1-\alpha)$ by using $\Omega_{\rm
m0}+\Omega_{\rm r0}+\Omega_{\rm X0}=1$. In next section, we will use
this result to constrain the parameters of SRDE model with
observational data.

\section{Observational constraints and the evolution of the SRDE}
Assuming that the Universe is spatially flat and taking $\Omega_{\rm
r0}=8\times 10^{-5}$, we constrain the parameters of SRDE model by
using the latest observational data including the Union+CFA3 sample
of 397 SNIa and the baryon acoustic oscillation (BAO) measurement
from the Sloan Digital Sky Survey (SDSS).

The SNIa data which provide the main evidence for the existence of
dark energy in the framework of standard cosmology. Here we consider
the latest published 397 Union+CFA3 SNIa data compiled in Table 1 of
\cite{Hicken}. This dataset add the CFA3 sample to the 307 Union
sample \cite{Kowalski}. Recently, Ref. \cite {gong} used these data
to study the accelerated expansion of Universe. The resulting
theoretical distance modulus $\mu_{\rm th}(z)$ is defined as
\begin{eqnarray}
\label{20}\mu_{\rm th}(z)\equiv 5\log_{10}D_{\rm L}(z)+\mu_0,
\end{eqnarray}
where $\mu_0\equiv 5\log_{10}h-42.38$ is the nuisance parameter
which can be marginalized over \cite{per05,Nesseris05}.

In order to break the degeneracies among the parameters, we consider
another observational constraint closely related the measurements of
the baryon acoustic oscillation peak in the distribution of SDSS
luminous red galaxies (LRG), the $A$ parameter, defined as
\begin{eqnarray}
A&\equiv&\Omega^{1/2}_{\rm
m0}(z^2_{1}E(z_1))^{-1/3}\left(\int^{z_1}_0dz/E(z)\right)^{2/3},
\end{eqnarray}
where $z_1=0.35$ is the effective redshift of the LRG sample. The
SDSS BAO measurement \cite{Eisenstein} gives $A_{\rm
obs}=0.469\,(n_s/0.98)^{-0.35}\pm 0.017$, where the scalar spectral
index is taken to be $n_s=0.960$ measured by WMAP5 \cite{Komatsu}.
Note that $A$ is independent of $H_0$, thus these quantities can
provide robust constraint as complement to SNIa data on SRDE.

As usual, assuming the measurement errors are Gaussian, the
likelihood function is ${\cal{L } } \propto e^{-\chi^2/2}$. The
model parameters yielding a minimal $\chi^{2}$ and a maximal
${\cal{L } }$ will be favored by the observations. Since the SNIa
and BAO are effectively independent measurements, we can simply
minimize their total $\chi^{2}$ value given by
\begin{eqnarray}
\label{21}\chi^2(\Omega_{\rm m0}, H_0)=\chi^{2}_{\rm
SNIa}+\chi^{2}_{\rm A},
\end{eqnarray}
where
\begin{eqnarray}
\chi^{2}_{\rm SNIa}&=&\sum^{N}_{i=1}\frac{(\mu^{\rm obs}_{\rm
L}(z_i)-\mu^{\rm th}_{\rm L}(z_i))^2} {\sigma^2_i},\\
\chi^{2}_{\rm A}&=&\left(\frac{A-A_{\rm obs}}{\sigma^2_{\rm
A}}\right)^2,
\end{eqnarray}
in order to find the best fit values of the parameters of the SRDE.
We obtain the best fit values of the parameters at $68\%$ confidence
level: $\Omega_{\rm m0}=0.259\pm0.016$ and $\alpha=0.261\pm0.0122$
with $\chi^2_{\rm min}=470.007$ (dof$=1.187$). For comparing, we
also constrain $\Lambda$CDM and RDE with the same data, and find :
$\Omega_{\rm m0}=0.285\pm0.021$ with $\chi^2_{\rm min}=465.719$
(dof$=1.173$) at $68\%$ confidence level for $\Lambda$CDM,
$\Omega_{\rm m0}=0.274\pm0.02$ and $\alpha=0.439\pm0.025$ with
$\chi^2_{\rm min}=466.156$ (dof$=1.174$) at $68\%$ confidence level
for RDE.

Recently, The best-fit values of parameters of RDE at $68.3\%$
confidence level are constrained to be:
$\Omega_{m0}=0.324^{+0.024}_{-0.022}$ and
$\alpha=0.371^{+0.023}_{-0.023}$, corresponding to
$\chi^2_{min}=483.130$, from the latest observational data including
the Union+CFA3 sample of 397 SNIa, the shift parameter of the cosmic
microwave background given by the five-year WMAP5 observations, and
BAO measurement \cite{Li2009}.

\begin{figure}
\includegraphics[width=8.5cm]{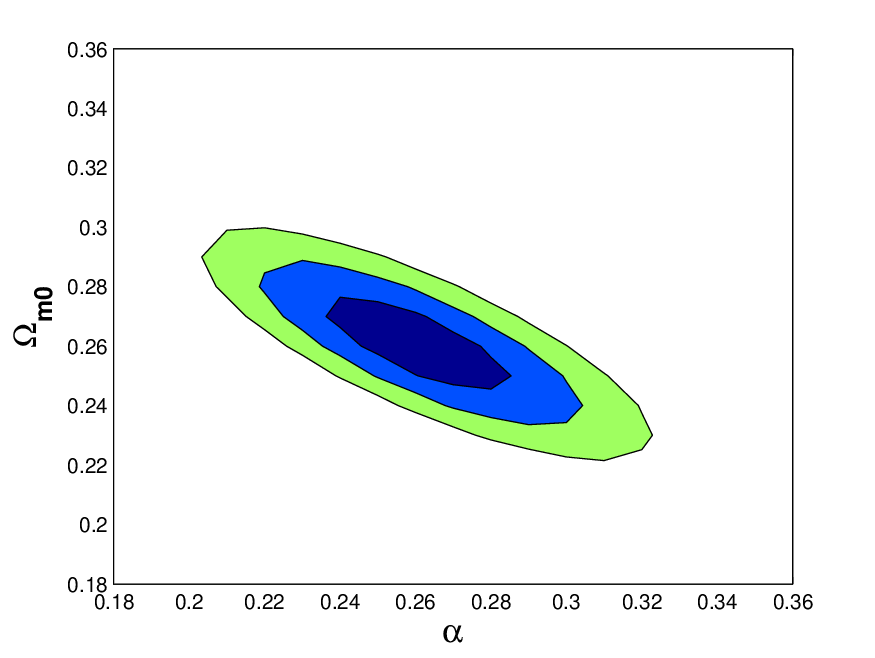}
\caption{The $68.3\%$, $95.4\%$ and $99.7\%$ confidence regions in
the $\alpha$-$\Omega_{\rm m0}$ plane for SRDE. \label{Fig1}}
\end{figure}

Figures \ref{Fig1} shows the $68.3\%$, $95.4\%$ and $99.7\%$ joint
confidence contours in the $\alpha$-$\Omega_{\rm m0}$ plane for
SRDE.
\begin{figure}
\includegraphics[width=8.5cm]{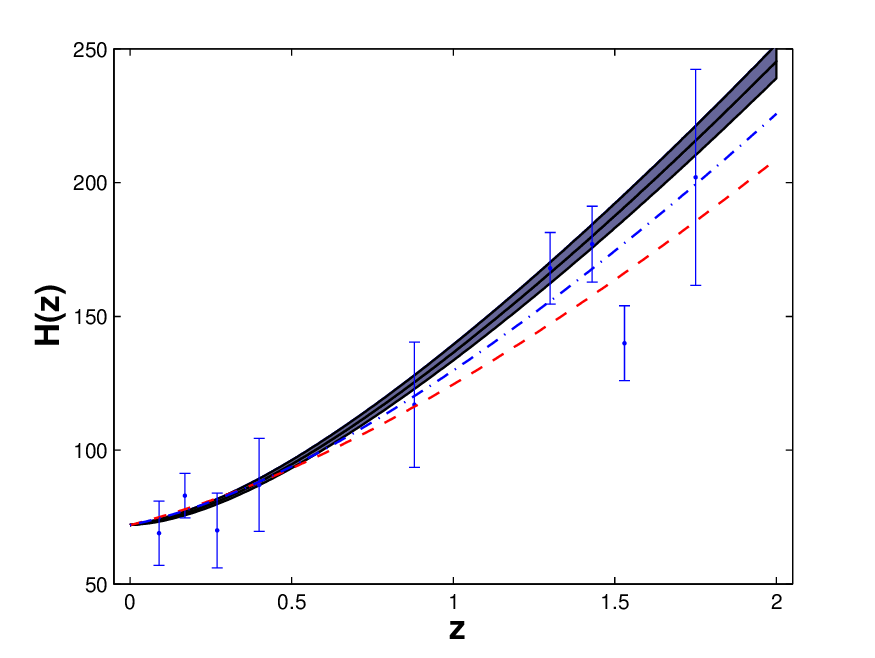}
\caption{The predicted values of the Hubble parameter $H$ of SRDE in
$68.3\%$ confidence level limits, compared with the observational
$H(z)$ data, $\Lambda$CDM (the dash line) and RDE (the dash-dot
line). \label{Fig4}}
\end{figure}
\begin{figure}
\includegraphics[width=8.5cm]{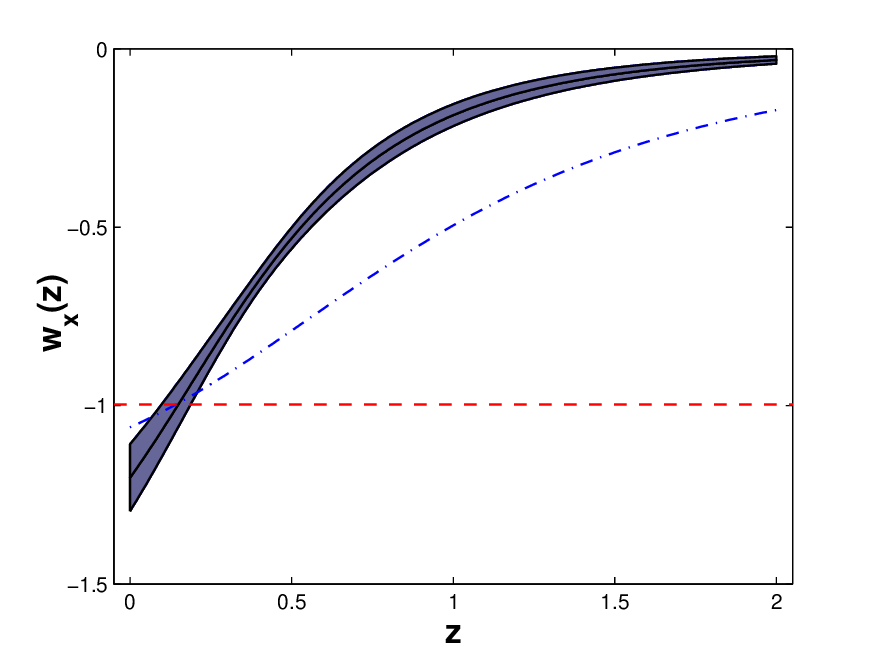}
\caption{The evolution of the equation of state parameter of SRDE in
$68.3\%$ confidence level limits, compared with $\Lambda$CDM (the
dash line) and RDE (the dash-dot line). \label{Fig2}}
\end{figure}
\begin{figure}
\includegraphics[width=8.5cm]{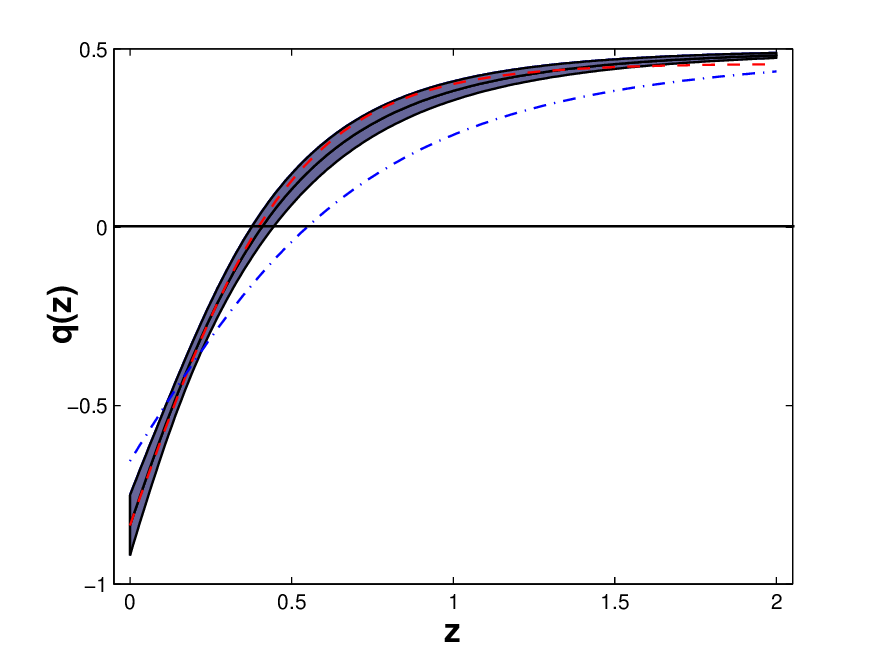}
\caption{The deceleration parameter as a function of redshift in
$68.3\%$ confidence level limits for SRDE, compared with
$\Lambda$CDM (the dash line) and RDE (the dash-dot line).
\label{Fig3}}
\end{figure}

The predicted values of the Hubble parameter $H$ of the SRDE in
$68.3\%$ confidence level limits compared with the observational
$H(z)$ data \cite{r8, r9, r12} is shown in figure \ref{Fig4}; the
$\Lambda$CDM and RDE cases are also presented for comparison. Here
we take $H_0=72$ kms$^{-1}$Mpc$^{-1}$ to calculate the Hubble
parameter $H$ of the SRDE, $\Lambda$CDM and RDE.

For $z\gtrsim 2$, the EoS parameter of SRDE runs closely to $-0$
shown in Fig. \ref{Fig2}, meaning the negative pressure of the SRDE
dark energy approaches to zero rapidly, compared with the cases of
RDE. The EoSs of SRDE and RDE cross $-1$ at $z\simeq0.14$, and
approach $w_{\rm X}\simeq-1.2$ and $w_{\rm X}\simeq-1.05$ at $z=0$,
respectively, consistent with the results obtained recently in Ref.
\cite{Enrique}. The properties in SRDE and RDE is interesting and
worthy to further study.

In Fig. \ref{Fig3}, we plot the evolution of de deceleration
parameter,
\begin{eqnarray}
q\equiv-1+\frac{1+z}{E}\frac{dE}{dz},
\end{eqnarray}
The present values of the deceleration parameter $q(z)$ for SRDE and
$\Lambda$CDM are found to be $q_{z=0}\sim -0.85$, while $q_{z=0}\sim
-0.7$ for RDE. The phase transition from deceleration to
acceleration of the Universe for SRDE and $\Lambda$CDM occur at the
redshift $z_{q=0}\sim 0.4$, comparable with that estimated from 157
gold data ($z_{\rm t}\simeq0.46\pm0.13$) \cite{Rie04}, while
$z_{q=0}\sim 0.6$ for RDE.

\section{structure formation}
It is a crucial test for any model of dark energy to study
perturbations in the dark matter/baryon component grow during the
matter-dominated era. Considering unified models of dark matter and
dark energy, such as Chaplygin gas, may suffer unstability
\cite{Bean2003, Sandvik, Perrotta, Amendola05, Zimdahl, Avelino},
although SRDE is not a unified model of dark matter and dark energy,
it behaves like dust in matter-dominated era, so it is important to
study whether this characteristic would affect the growth rate of
dark matter/baryon density perturbation and upset the usual
structure formation scenario.

\begin{figure}
\includegraphics[width=8.5cm]{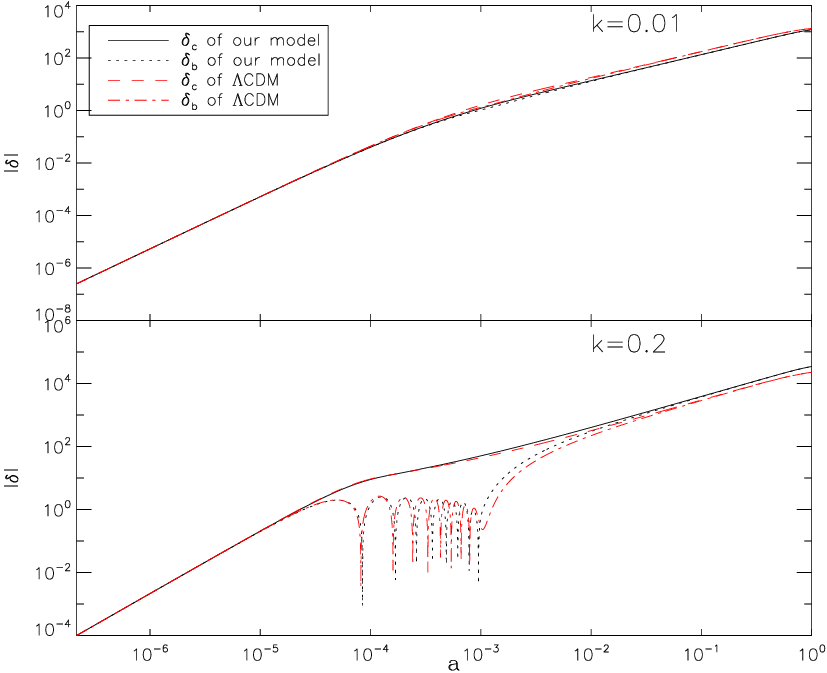}
\caption{Evolution of the dark matter/baryon density perturbations
in the SRDE and $\Lambda$CDM model for two wavenumbers: $k=0.01$
(upper panel) and 0.2 h Mpc${}^{-1}$ (lower panel). $\delta_c$ and
$\delta_b$ are the perturbations of cold dark matter and baryon
respectively. \label{Fig5}}
\end{figure}

In Fig. \ref{Fig5}, we compare the evolution of dark matter/baryon
density perturbations in SRDE with that in the $\Lambda$CDM model by
a generated multifunctional module of Boltzmann code \cite{Wu,Wu1},
which originally based on CAMB \cite{camb}. The parameters is the
best-fit values as discussed earlier: $\Omega_{\rm k0}=0$,
$\Omega_{\rm m0}=0.259$ ($\Omega_{\rm b}=0.05$, $\Omega_{\rm
c}=0.204$), and $\alpha=0.261$. As shown in Fig. \ref{Fig5}, the
differences between SRDE and the $\Lambda$CDM model are very small,
almost invisible for the large scale perturbations (e.g. $k=0.01$ h
Mpc${}^{-1}$ modes); For the small scale perturbations (e.g. $k=0.2$
h Mpc${}^{-1}$ modes), the amplitudes of $\delta_{\rm c}$ and
$\delta_{\rm b}$ in SRDE are slightly larger than these in $\Lambda
\textrm{CDM}$ model, due to the extra dust-like component in SRDE.

\begin{figure}
\includegraphics[width=8.5cm]{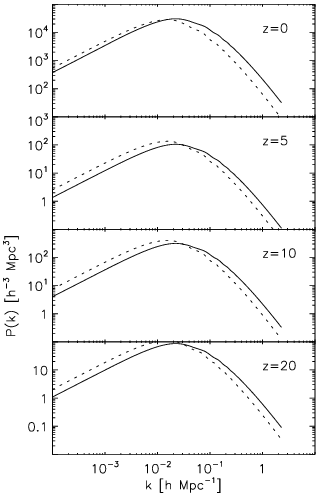}
\caption{The matter power spectra at different redshifts. From top
to bottom: $z=0$, $z=5$, $z=10$, and $z=20$. The solid and dashed
curves represent SRDE and $\Lambda \textrm{CDM}$ model,
respectively. \label{Fig6}}
\end{figure}

The matter power spectra at different redshifts is plotted in
Fig.~\ref{Fig6}. The turnover of matter power spectra occurs at
smaller scale compared with the $\Lambda$CDM model as showed in Fig.
~\ref{Fig6}, this is because the extra dust-like component appears
in SRDE, then in turn it will shift the matter-radiation equality
$a_{\rm eq}$ to small value.  From Fig.~\ref{Fig6}, we also can
conclude that the growth rate of SRDE is somewhat different from the
$\Lambda$CDM due to different ratios of matter power spectrum of
SRDE to that of the $\Lambda$CDM model at different redshifts.
However, we should say that the deviation of the shapes of the
spectra from $\Lambda$CDM model is small, and it is expected to be
fitted well with the observation by adjusting other parameters, such
as $\sigma_8$ and $n_S$.

\begin{figure}
\includegraphics[width=8.5cm]{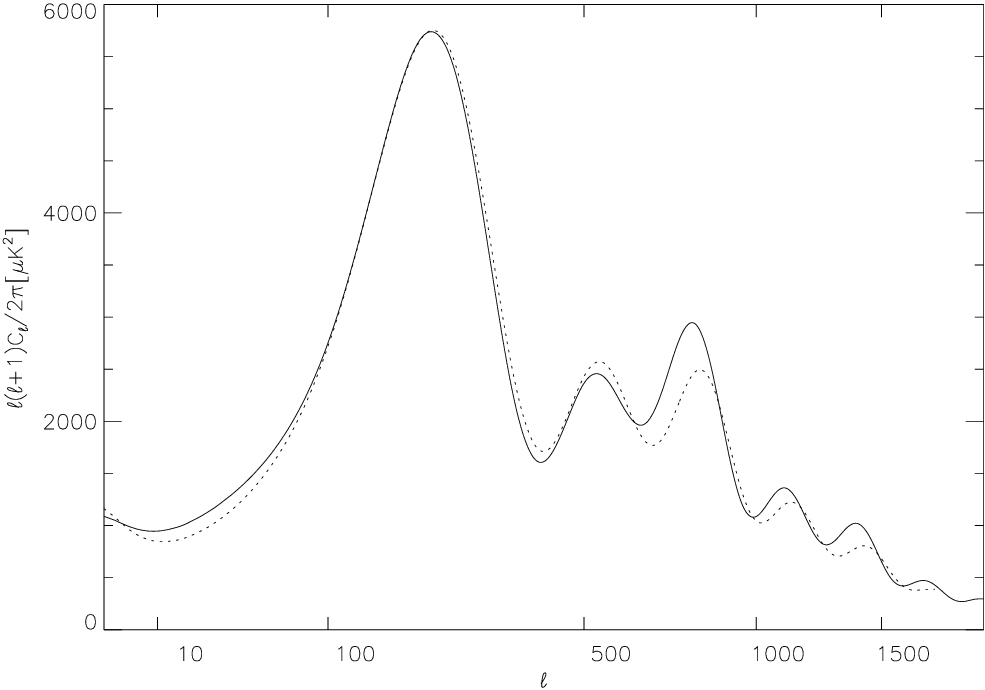}
\caption{The theoretical CMB temperature angular power spectrum of
SRDE (solid curve) compared with  $\Lambda \textrm{CDM}$ model
(dashed curve). We normalize two amplitudes of spectra to same at
the first peak. \label{Fig7}}
\end{figure}

To compare with the CMB anisotropy of $\Lambda$CDM model, we present
CMB angular power spectra of SRDE in Fig. \ref{Fig7}. The plot shows
that two results are consistent with each other well on the large
scale. On the small scale, the difference is still in acceptable
level due to large error bars in this range. On the other hand, this
mean we could test SRDE from small scale data, for example, Planck
data in the future. These kinds of data could be helpful to
distinguish SRDE from $\Lambda$CDM model.

\section{conclusions and discussions}
In this article, we have shown that replace Ricci scalar curvature
in RDE model with spatial Ricci scalar curvature, $\rho_X\propto
R_{\rm s}$, the resulting SRDE model is viable phenomenologically.
The dark energy component in SRDE contains a term evolves like
nonrelativistic matter, a term increases with decreasing redshift,
and a term serves as dark radiation which RDE doesn't have. Like
RDE, SRDE avoided the casuality problem of holographic dark energy,
because the dark energy is determined by the locally determined
spatial Ricci scalar curvature, not the future event horizon.
Because SRDE model is not associated Planck or other high energy
physics scale, but with the size of space-time curvature, the fine
tuning problem can be avoided.

Assuming that the Universe is spatially flat, we have placed
observational constraints on SRDE scenario with the Union+CFA3
sample of 397 SNIa and the BAO measurement from the SDSS. For SRDE,
we have obtained the best fit values of the parameters at $68\%$
confidence level: $\Omega_{\rm m0}=0.259\pm0.016$ and
$\alpha=0.261\pm0.0122$ with $\chi^2_{\rm min}=470.007$
(dof$=1.187$). Withe these values of parameters, we have studied
evolutions of Hubble parameters, parameters of EoS and the
deceleration parameters. SRDE evolves like nonrelativistic matter
when $z\longrightarrow2$, while RDE not. The present values of the
deceleration parameter $q(z)$ for SRDE is found to be $q_{z=0}\sim
-0.85$. The phase transition from deceleration to acceleration of
the Universe for SRDE occurs at the redshift $z_{q=0}\sim 0.4$,
comparable with that estimated from 157 gold data ($z_{\rm
t}\simeq0.46\pm0.13$) \cite{Rie04}. We found the equation of state
of SRDE crosses $-1$ at $z\simeq-0.14$.

We have discussed the dark matter/baryon density perturbations, the
differences between SRDE and the $\Lambda$CDM model are very small
for the large scale perturbations, while a little larger for the
small scale perturbations. As expected, due to the extra dust-like
component, the growth rate of SRDE differs from that in $\Lambda$CDM
model, and the matter-radiation equality occurred at smaller $a_{\rm
eq}$. We also have shown that the CMB angular power spectra of SRDE
is consistent with that of $\Lambda$CDM on the large scale, while
difference slightly on the small scale. This mean we could test SRDE
from small scale data.

The properties of SRDE, the presented dark radiation like as in
brane dark energy and in Ho\u{r}ava-Lifshitz cosmology and the
behavior crossing $-1$ like as in quintom, may be interesting in
future study. We note, however, we have been inspired by the
holographic principle when constructing SRDE, SRDE does not
necessarily have to be connected with the holographic principle.
Although SRDE is phenomenologically successful and theoretically
interesting, its physical mechanism is await further study.

\begin{acknowledgments}
This study is supported in part by Research Fund for Doctoral
Programs of Hebei University No. 2009-155, and by Open Research
Topics Fund of Key Laboratory of Particle Astrophysics, Institute of
High Energy Physics, Chinese Academy of Sciences, No.
0529410T41-200901.
\end{acknowledgments}

\bibliography{apssamp}

\end{document}